\documentstyle[prl,tighten,multicol,aps,epsfig]{revtex}
\begin{document}
%\draft
\title
{Polymer confinement in undulated membrane boxes and tubes}
\author { Tomonari Dotera}
\address{ Saitama Study Center, the University of the Air,
682-2 Nishiki-cho, Omiya 331-0851, Japan }
\author {Yasuo Y. Suzuki }
\address{NTT Basic Research Laboratories, Atsugi 243-0198, Japan}
\date{\today}
\maketitle
\begin{abstract}
We consider quantum particle or Gaussian polymer confinement
between two surfaces and
in cylinders with sinusoidal undulations.
In terms of the variational method,
we show that the quantum mechanical wave equations
have lower ground state energy in these geometries
under long wavelength undulations,
where bulges are formed and waves are localized in the bulges.
It turns out correspondingly that
Gaussian polymer chains in undulated boxes
or tubes acquire higher entropy
than in exactly flat or straight ones.
These phenomena are explained
by the uncertainty principle for quantum particles,
and by a {\it polymer confinement rule} for Gaussian polymers.
If membrane boxes or tubes are flexible,
polymer-induced undulation instability is suggested.
We find that the wavelength of undulations at the threshold of instability
for a membrane box is almost twice the distance
between two walls of the box.
Surprisingly we find that the instability for tubes
begins with a shorter wavelength
compared to the ``Rayleigh'' area-minimizing instability.
\end{abstract}
\pacs{PACS numbers: 61.25.Hq, 87.22.Bt, 47.20.Ma}

\begin{multicols}{2}
\narrowtext

\section{Introduction}
Recently much attention has been paid to the structure
and dynamics of polymer chains restricted by two surfaces,
or restricted in cylindrical pores \cite{Ca,Daoud,BC,Lal}.
These conditions are relevant to a broad class of
applications and biological functions, such as filtration, gel
permeation chromatography, heterogeneous catalysis,
oil recuperation.

When a polymer is confined to a smaller space,
the entropy loss of a chain is higher.
Here we call this simple rule the {\it polymer confinement rule}.
It has been employed casually in the literature, however, it has
a close connection with the uncertainty principle in quantum mechanics.
Indeed, on a theoretical side, quantum mechanical wave equations
have been applied to elucidate the behavior of the Gaussian chains
when the radius of gyration of a chain is much larger than these structures.
Therefore, the solutions of wave equations in some geometries are
useful both for mesoscopic quantum physics and polymer physics.

The most prominent recent example is the calculation for
curved geometries.
Goldstone and Jaffe \cite{GJ} have shown that
the bend of two and three dimensional tubes with a constant cross section
lowers the ground state energy of a quantum particle constrained
in the tubes.
Correspondingly, the cylindrical bend of two parallel walls (2D tube)
with a constant width reduces the entropy loss \cite{YPSW}.
Yaman {\it et~al.} have shown in their series of works
that entropic interactions between curved membranes and macromolecules
such as flexible chain molecules and rigid rods may change the bare elastic
constants of single membranes and bilayers \cite{Safran}.

In this paper, we focus on tubes in two and three dimensions with
sinusoidal undulations keeping tube volumes constant.
This lowers the ground state energy of quantum particles,
and thus raises the entropy of Gaussian polymer chains.
Corrugated walls induce additional kinetic energy along the walls,
however,
if bulges are formed, the waves are less confined in the transverse
direction and localized in the bulges, which may decrease
the total energy because of the uncertainty principle.
In the same way,  according to the polymer confinement rule,
polymer chains may favor the undulations of tubes.

In section II, we first describe the polymer confinement rule,
an intuitive argument for why a polymer in a confined
space loses entropy.
This approach is microscopic, which is different from 
the well-known scaling argument\cite{Daoud}.
Second, we provide the calculation procedure for Gaussian chains
using quantum mechanics which is given in Ref.\cite{Ca}.
In section III,
we elucidate the effect of undulations
using the variational method.
We show that the wave equations
have lower energy under long wavelength undulations.
Correspondingly, long Gaussian chains acquire higher entropy
with undulating constraints than with exactly flat or straight ones.
For the three dimensional case, we compare the undulation effect
with the Rayleigh area-minimizing instability \cite{Safran,DWT}.
In the final section, we discuss implications of
polymer-mediated entropic force for membrane boxes and tubes.

\section{Correspondence to quantum mechanics}

\subsection{The polymer confinement rule}
The entropy reduction of a confined random walk of $N$ steps
between two parallel flat walls (distance $\Delta x$ )
can be derived intuitively through the following microscopic reasoning.
The number of steps that span the distance between the walls
is described by
$\Delta x\sim l {N'}^{1/2}$, and the gyration radius
in a free space $R_{\rm g}(\,\gg\Delta x)$ is
expressed as $R_{\rm g}\sim l N^{1/2}$,
where $l$ is a step length.
Here $\sim$ implies that the numerical factors have been ignored.

Steps touching the walls should return.
This requirement is the source of the entropy reduction \cite{Unpub}.
The chain reflects off the walls about $N/N'$ times.
Let $z$ be the number of nearest positions on a lattice for a step.
Then the number $W$ of total configurations
of the confined chain is represented by
\begin{equation}
 W  \sim  z^{N-N/N'} (z/2)^{N/N'}
=z^N(1/2)^{N/N'},
\end{equation}
where the combination factor has been omitted.
Then the entropy loss due to the walls for the Gaussian chain is
$\Delta S = k_{\rm B} N/N' \ln 2 \sim k_{\rm B} R_{\rm g}^2/(\Delta x)^2$.
We write
\begin{equation}
(\Delta x)^2 \Delta S \sim k_{\rm B} R_{\rm g}^2,
\label{pqr}
\end{equation}
which corresponds to the uncertainty principle:
\begin{equation}
(\Delta x)^2 \Delta E \sim (\Delta x)^2 (\Delta p)^2 /m \sim h^2 /m,
\end{equation}
where $E$, $x$, $p$, and $m$ are the energy, position, momentum and mass
of a quantum particle, and $h$ is the Planck constant.
On this basis, one can say that a polymer tends to escape
from narrower spaces into wider or open spaces; in other words,
the polymer tends to localize in wider spaces
of the confined geometries, the same as with quantum particles.

\subsection{Analytical theory for Gaussian chains}
For a random walk, we consider the total number of paths
that connect $r$ and $r'$ with $N$ steps, $z^N G({\bf r',r},N)$,
where $z$ is the number of neighboring sites.
The boundary condition is
\begin{equation}
G({\bf r',r},0)= \delta_{\bf r',r}.
\label{bdcond}
\end{equation}
It is easy to show $G(r',r,N)$ is the solution of the diffusion equation:
\begin{equation}
\left( \frac{\partial}{\partial N}
- \frac{l^2}{6} \nabla_{\bf r}^2  \right) G({\bf r',r},N)=0,
\end{equation}
where $l$ is the lattice constant, or the step length.

With the eigenfunction expansion:
\begin{equation}
G({\bf r',r},N)=l^3 \sum_n \Psi_n^*({\bf r'}) \Psi_n({\bf r})
\exp\left(-\frac{N l^2 E_n}{6}\right),
\label{G}
\end{equation}
then, the problem reduces to the wave equation,
\begin{equation}
\nabla^2 \Psi_n({\bf r})+E_n\Psi_n({\bf r})=0
\label{wave}
\end{equation}
with the boundary conditions.
As is known, the eigenfunctions satisfy the orthogonality
and completeness conditions:
\begin{eqnarray}
\int dr \Psi_m^*({\bf r}) \Psi_n({\bf r}) = \delta_{mn},
\\
\sum_n \Psi_n^*({\bf r'}) \Psi_n({\bf r}) = \delta({\bf r-r'}).
\end{eqnarray}
Since the continuous limit is
\begin{equation}
\lim_{l\rightarrow0} \delta_{\bf r',r}/l^3 = \delta({\bf r-r'}),
\end{equation}
Eq.(\ref{G}) satisfies the boundary condition Eq.(\ref{bdcond}).

For excited states ($i>1$), when an equality
\begin{equation}
R_{\rm g}^2 \ ( E_i - E_1 ) \gg 1,
\label{gsd}
\end{equation}
is satisfied, the situation is called ground state dominance,
which is our interest in this paper.
Then we have,
\begin{equation}
G({\bf r,r'},N) \approx  \Psi_0^*({\bf r'}) \Psi_0({\bf r})
\exp\left(-\frac{N l^2 E_1}{6}\right).
\end{equation}

The partition function is a sum over all configurations:
\begin{equation}
Z=\frac{1}{V}\int d{\bf r} d{\bf r}' G({\bf r,r'},N).
\end{equation}
Integrating out all uninteresting degrees of freedom, and we obtain
the main term in the associated entropy change,
\begin{equation}
S=k_{\rm B} \ln Z \approx -k_{\rm B} \frac{N l^2 E_1}{6}.
\end{equation}

For a Gaussian chain between two flat walls,
the discrete part of the eigenvalues of Eq.(\ref{wave}) is
\begin{equation}
E_n=\frac{n^2\pi^2}{(\Delta x)^2},\quad n=1,2,3,\cdots.
\end{equation}
To obtain the main term, we ignore the continuous part of
the eigenvalues associated with two directions along walls.
Then Eq.(\ref{gsd}) is fulfilled.
Thus, we have
\begin{equation}
S \approx -k_{\rm B} \frac{\pi^2}{6} \frac{R_{\rm g}^2 }{(\Delta x)^2}.
\end{equation}
This is consistent with Eq.(\ref{pqr}).

\section{Variational proofs}
Let $C$ be a tube in two dimensions (box) or in three dimensions.
We will consider the wave equation,
\begin{equation}
(\nabla^2+E)\psi({\bf r})=0,
\end{equation}
in $C$ subjected to the Dirichlet condition on walls: $\psi(\partial C)=0$.
The following calculations are applicable to both quantum particles and
Gaussian polymers.

We define
\begin{eqnarray}
\sigma[\psi] \equiv \left( \int_{C}d^D{\bf r} \psi \nabla^2 \psi \right)
\cdot
\left( \int_{C}d^D{\bf r} \psi^2 \right)^{-1},
\label{E}
\end{eqnarray}
where $D$ is the dimension of the system.
Our aim is to show
\begin{equation}
\Delta[\psi]\equiv\sigma[\psi] - \sigma_0>0,
\end{equation}
where $\sigma_0$ corresponds to the ground state
for the exactly flat or straight case.
Since perturbations are sinusoidal, the integrals can be limited
to a single wavelength, that of the perturbations.
We compute a critical wavelength,
down to which $\Delta[\psi]$ is positive.
Our evaluation gives an upper bound for the critical wavelength, 
because we employ a trial function
for $\psi$ in Eq.(\ref{E}).

\subsection{In two dimensions}
As shown in Fig.1,
a box is defined between two parallel lines described by
\begin{equation}
y_{\pm}(x)=\pm \frac{d}{2}(1+\epsilon \sin ax), \quad 0<\epsilon < 1
\end{equation}
with a width undulation function $w(x)$:
\begin{equation}
w(x) = d(1+\epsilon \sin ax),
\end{equation}
where $ a = 2\pi/\lambda,$ $\lambda$ is the period of
the undulation, and $d$ is the mean distance.
Note that the volume is kept constant.

\begin{figure}
\epsfysize=3.5cm
\centerline{\epsfbox{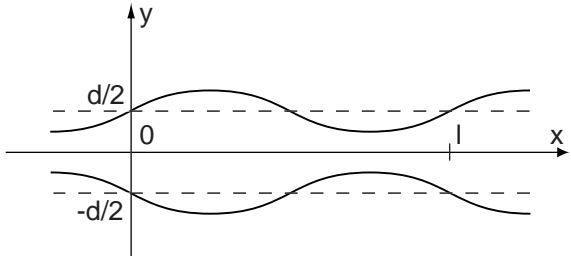}}
\caption{Undulated box in two dimensions: $\lambda$ is the period of
the undulation, and $d$ is the mean distance between walls.}
\label{fig1}
\end{figure}

We will prove, with the variational method, that long-wave undulation
lowers the total energy.
Choose a candidate function with a variational constant $\delta$:
\begin{equation}
\psi(x,y)=\cos \left(\frac{\pi y}{w}\right)
(1+\delta \epsilon \sin ax),
\quad 0 \leq \delta \leq 1/\epsilon
.
\end{equation}
$\delta>0$ implies localization at bulges.
The denominator of Eq.(\ref{E}) is
\begin{eqnarray}
\frac{1}{d}
\int_{-\frac{\lambda}{2}}^{\frac{\lambda}{2}}
dx
\int_{-\frac{w}{2}}^{\frac{w}{2}}
dy
\psi^2
=\frac{\lambda \gamma}{2}
,
\end{eqnarray}
where
\begin{equation}
\gamma = 1+\left(\delta+\frac{\delta^2}{2} \right)\, \epsilon^2>0.
\end{equation}

It is elementary to calculate
\begin{eqnarray}
\frac{\partial ^2 \psi}{\partial y^2} &=&
-\left(\frac{\pi}{w}\right)^2
\cos \left(\frac{\pi y}{w}\right)
(1+\delta \epsilon \sin ax)
,
\\
\frac{\partial ^2 \psi}{\partial x^2}  &=&
\left[
\pi\frac{w''}{w^2}y\sin \left(\frac{\pi y}{w}\right)
-2\pi\frac{w'^2}{w^3}y\sin \left(\frac{\pi y}{w}\right)
\right.
\nonumber
\\
&-&
\left.
\left(\pi\frac{w'}{w^2}\right)^2y^2\cos \left(\frac{\pi y}{w}\right)
\right]
(1+\delta \epsilon \sin ax)
\nonumber
\\
&+&2\delta \epsilon a \pi \frac{w'}{w^2}
y \sin \left(\frac{\pi y}{w}\right)
\cos ax
\\
&-&\delta \epsilon a^2
\cos \left(\frac{\pi y}{w}\right)
\sin ax
.
\nonumber
\end{eqnarray}
Then, a brief calculation yields
\begin{eqnarray}
\label{A1}
\frac{1}{d}\int dy\psi\frac{\partial ^2 \psi}{\partial y^2} &=&
-\frac{\pi^2}{2d^2}
\frac{(1+\delta \epsilon \sin ax)^2 }{1+\epsilon \sin ax}
,
\end{eqnarray}
\begin{eqnarray}
\label{A2}
\frac{1}{d}\int dy\psi\frac{\partial ^2 \psi}{\partial x^2} &=&
-
\frac{\pi^2+6}{24}
\frac{\epsilon^2 a^2 \cos^2 ax}{1+\epsilon \sin ax}
(1+\delta \epsilon \sin ax)^2
\nonumber
\\
&-& \left(\frac{\delta^2}{2} + \delta \right)
\epsilon^2 a^2 \sin^2ax
\\
&+&\frac{\delta \epsilon^2 a^2}{2} \cos^2ax
+\frac{\delta^2 \epsilon^3 a^2}{2} \cos^2ax \sin ax
\nonumber
\\
&-&\frac{3 \delta^2 \epsilon^3 a^2}{4} \sin^3ax
-\frac{(\epsilon a^2+2\delta \epsilon a^2)}{4} \sin ax
.
\nonumber
\end{eqnarray}
Using formulae given in Appendix A, we obtain
\begin{eqnarray}
\gamma
\sigma_y[\psi]&=&
-\frac{\pi^2}{d^2}
\left(
\frac{1}{\sqrt{1-\epsilon^2}}
- 2\delta A + \delta^2 A
\right),
\label{plus}
\\
\gamma
\sigma_x[\psi]&=&
-a^2 B D
-
2\delta a^2
\left( \frac{\epsilon^2}{4} -CD
\right)
\\
&&
-
\delta^2 a^2
\left( \frac{\epsilon^2}{2} + CD
\right),
\nonumber
\end{eqnarray}
where $\sigma[\psi]=\sigma_y[\psi]+\sigma_x[\psi]$, and where
\begin{eqnarray}
A &=& \frac{1}{\sqrt{1-\epsilon^2}}-1
=\frac{\epsilon^2}{2}+\frac{3\epsilon^4}{8}+\frac{5\epsilon^6}{16}+\cdots ,
\label{ABC}
\\
B &=& 1- \sqrt{1-\epsilon^2}
=\frac{\epsilon^2}{2}+\frac{\epsilon^4}{8}+\frac{\epsilon^6}{16}+\cdots ,
\\
C &=& B - \frac{\epsilon^2}{2}
=\frac{\epsilon^4}{8}+\frac{\epsilon^6}{16}+\cdots ,
\label{ABCD}
\\
D &=& \frac{\pi^2+6}{12}.
\end{eqnarray}

For $\epsilon > 0$,
we evaluate the deviation $\Delta[\psi]$ for $\sigma_0 = -\pi^2/d^2$.
To show $\Delta[\psi]$ is positive, we write as
\begin{eqnarray}
\gamma \Delta[\psi]&=&
-\frac{\pi^2}{d^2}A - \frac{\pi^2}{\lambda^2} 4BD
\nonumber
\\
&+& 2\delta
\left[ \frac{\pi^2}{d^2} \left(A +\frac{\epsilon^2}{2} \right)
- \frac{\pi^2}{\lambda^2} ( -4CD +\epsilon^2 ) \right]
\\
&-&\delta^2
\left[ \frac{\pi^2}{d^2} \left(A -\frac{\epsilon^2}{2} \right)
+ \frac{\pi^2}{\lambda^2} ( 4CD + 2\epsilon^2 ) \right].
\nonumber
\end{eqnarray}
From this, it easy to show that there exists $\delta$ that
makes $\Delta[\psi]$  positive in the long-wave undulation limit:
$\lambda \rightarrow \infty$; $\delta=1$ for instance.
Therefore, the ground state energy should be lower than
that in the flat plane case.
Notice that as is seen in Eq.(\ref{plus})
the decrease in energy $2\delta A$ appears in the transverse direction.

In order to estimate the critical wavelength $\lambda_0$,
an approximation up to order $\epsilon^2$ is
\begin{eqnarray}
(\gamma /\epsilon^2) \Delta[\psi]&\approx&
-2 \delta^2 \left( \frac{\pi^2}{\lambda^2} \right)
+ 2\delta \left( \frac{\pi^2}{d^2} - \frac{\pi^2}{\lambda^2} \right)
\nonumber
\\
&&-\frac{\pi^2}{2d^2} - 2D \frac{\pi^2}{\lambda^2}.
\label{last1}
\end{eqnarray}
Maximizing $\Delta[\psi]$ by changing the variational constant $\delta$,
we compute the critical wavelength from $\Delta[\psi]=0$. Thus,
\begin{eqnarray}
\frac{\lambda_0}{d}=\left( \frac{3+\sqrt{\frac{4\pi^2}{3}+13}}{2}
\right)^{\frac{1}{2}}=2.014.
\end{eqnarray}
The critical wavelength is about two times the width of the box,
implying that the bulge is at least the size of
the width between walls.
It is a quite reasonable value, because, in terms of quantum mechanics,
if the undulation is shorter than $\lambda_0$, then
the kinetic energy along walls will be higher than the confinement energy.

\subsection{In three dimensions}
It was shown by Plateau, later pursued by Rayleigh,
then known as the Rayleigh instability,
that the undulation of a tube
with wavelength exceeding its circumferences
reduces the surface area.
Therefore, it is interesting to compare the
polymer-mediated instability and the Rayleigh instability.

Consider a tube of unperturbed radius $R_0$ and
an undulation function (Fig.2): 
\begin{equation}
R(z)=R_c(1+\epsilon \sin az), \quad 0<\epsilon < 1,
\end{equation}
where $a=2\pi/\lambda$, and $R_c$ is determined by the constant
volume condition of the tube. Thus,
\begin{equation}
R_c^2=R_0^2(1+\epsilon^2/2)^{-1}.
\label{rad}
\end{equation}

\begin{figure}
\epsfysize=3cm
\centerline{\epsfbox{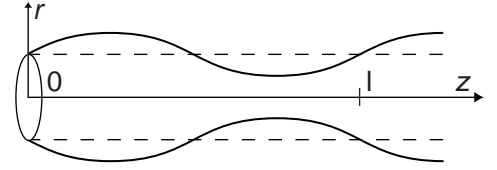}}
\caption{Undulated tube in three dimensions: $\lambda$ is the period of
the undulation.}
\label{fig2}
\end{figure}

In order to remind readers of the Rayleigh instability,
we calculate the surface area of the tube given by
\begin{equation}
A(\epsilon; \lambda) = 2\pi \int_0^\lambda
R \sqrt{1+\left(\frac{dR}{dz}\right)^2} dz.
\end{equation}
We expand the square root for small perturbations and use Eq.(\ref{rad}).
The area is
\begin{equation}
A(\epsilon; \lambda) \approx 2\pi R_0 \lambda
\left[1+\frac{\epsilon^2}{4}\left(\frac{(2\pi
R_0)^2}{\lambda^2}-1\right)\right].
\end{equation}
Therefore, if the wavelength is greater than the circumference of the tube,
the area is less than that of the unperturbed perfect cylinder.

Now, we will prove that a long-wave undulation lowers the total energy with the
variational method.
Choose a trial function with a variational constant $\delta$:
\begin{equation}
\psi(r,z)=J_0 \left(\frac{\alpha r}{R}\right)
(1+\delta \epsilon \sin az),
\quad 0 \leq \delta \leq 1 / \epsilon .
\end{equation}
where $J_0(r)$ is the 0-th Bessel function:
\begin{eqnarray}
\frac{d^2J_0\left(\frac{\alpha r}{R}\right)}{dr^2}
+\frac{1}{r}\frac{dJ_0\left(\frac{\alpha r}{R}\right)}{dr}
+\left(\frac{\alpha}{R}\right)^2J_0\left(\frac{\alpha r}{R}\right)=0,
\end{eqnarray}
and where $\alpha=2.40483$ is the smallest zero of $J_0(x)$.
The denominator of Eq.(\ref{E}) is
\begin{eqnarray}
\int_{-\frac{\lambda}{2}}^{\frac{\lambda}{2}}
dz
\int_{0}^{R}
dr 2\pi r
\psi^2
=\pi \lambda  R_c^2 \beta\ \gamma,
\end{eqnarray}
where $\beta=J_1^2(\alpha)$ and
\begin{equation}
\gamma = 1+\frac{1}{2}(\delta^2+4\delta+1)\,\epsilon^2
+\frac{3}{8}\delta^2\epsilon^4.
\end{equation}

It is elementary to calculate
\begin{eqnarray}
\frac{\partial ^2 \psi}{\partial r^2}
+\frac{1}{r}\frac{\partial  \psi}{\partial r}
&=&
-\left(\frac{\alpha}{R}\right)^2
J_0 \left(\frac{\alpha r}{R}\right)
(1+\delta \epsilon \sin az),
\end{eqnarray}
\begin{eqnarray}
\frac{\partial ^2 \psi}{\partial z^2}  &=&
\left[
-\alpha\frac{R''}{R^2}rJ_0' \left(\frac{\alpha r}{R}\right)
+\alpha\frac{R'^2}{R^3}r J_0' \left(\frac{\alpha r}{R}\right)
\right.
\nonumber
\\
&-&
\left.
\left(\alpha\frac{R'}{R^2}\right)^2r^2 J_0 \left(\frac{\alpha r}{R}\right)
\right]
(1+\delta \epsilon \sin az)
\nonumber
\\
&-&2 \alpha \delta \epsilon a \frac{R'}{R^2}r
J_0' \left(\frac{\alpha r}{R}\right)
\cos az
\\
&-&\delta \epsilon a^2
J_0 \left(\frac{\alpha r}{R}\right)
\sin az,
\nonumber
\end{eqnarray}
where, $J_0'(x) = dJ_0(x)/dx$, $R'=dR(z)/dz$, and $R''=d^2R(z)/dz^2$.

Using the calculation in Appendix B, we then obtain
\begin{eqnarray}
\gamma \sigma[\psi] &=& -\frac{\alpha^2(2+\delta^2 \epsilon^2)}{2 R_c^2}
\\
&-&a^2\left(\frac{4+\alpha^2}{6}\epsilon^2+\delta
\epsilon^2+\frac{\delta^2\epsilon^2 }{2}
+\frac{13+\alpha^2}{24}\delta^2\epsilon^4\right).
\nonumber
\end{eqnarray}
For $\epsilon > 0$,
we evaluate $\Delta[\psi]$ for $\sigma_0 = -\alpha^2/R_0^2=-b^2$.
\begin{eqnarray}
\frac{\gamma \Delta[\psi]}{\epsilon^2}
=&-&\left[
\frac{-b^2 \epsilon^2}{8}
+a^2
\left(
\frac{1}{2} + \frac{13+\alpha^2}{24}\epsilon^2 \right)
\right]\delta^2
\nonumber
\\
&+&(2b^2-a^2)\delta
-a^2 \frac{4+\alpha^2}{6}.
\end{eqnarray}

An approximation up to $\epsilon^2$ is
\begin{eqnarray}
\frac{\gamma \Delta[\psi]}{\epsilon^2}
\approx &-& \frac{a^2}{2} \, \delta^2
+(2b^2-a^2)\,\delta
-a^2 \frac{4+\alpha^2}{6}.
\label{last2}
\end{eqnarray}
Maximizing $\Delta[\psi]$ by varying $\delta$,
we calculate the critical wavelength from $\Delta[\psi]=0$.
Thus, we have
\begin{eqnarray}
4\alpha^4 x^4 -4\alpha^2 x^2 -\frac{1+\alpha^2}{3}=0,
\end{eqnarray}
where
\begin{equation}
x \equiv \frac{\lambda}{2\pi R_0}=\frac{b}{\alpha a},
\end{equation}
is the ratio of the critical wavelength to that of the Rayleigh instability.
Then the minimum $x_0>0$ is
\begin{equation}
x_0 = \left( \frac{1+\sqrt{\frac{4+\alpha^2}{3}}}{2\alpha^2} \right)^{1
\over 2}
= 0.493.
\end{equation}
The critical wavelength is shorter than that of the Rayleigh instability.

\section{Discussion}
We have shown that quantum mechanical particles confined in
undulated boxes or tubes have lower energy, when the
wavelength is greater than certain values,
comparable to the width between walls or the radius of tubes.
We have explained that the effect can easily be interpreted
by the uncertainty principle.
Quantum mechanical calculations immediately imply that
long Gaussian chains in undulated boxes or
tubes acquire higher entropy than in exactly flat or straight ones.
Furthermore, it can be explained
by the polymer confinement rule established in this paper,
which is quite analogous to the uncertainty principle.

This polymer-mediated entropic force
may play an important role in a number of systems:
polymers in cell membranes, vesicles,
microemulsions, and polymers confined in lamellar or cylindrical phases
of surfactant and homopolymers blending into those of block copolymer systems.
For instance, deformable flat membrane boxes
with constant width $d$
containing the Gaussian chains ($R_{\rm g} \gg d$)
are unstable against the undulations.
In the same way, cylindrical tubes with radius $R$
containing the Gaussian chains ($R_{\rm g} \gg R$)
are unstable against the undulations.
From Eqs.(\ref{last1}) and (\ref{last2}), it is easy to see
that the entropy gain increases with increasing  
undulation amplitude. 
Therefore, the undulation amplitude will grow 
until it is balanced by the elastic restoring force of the membrane, 
or until the membrane ruptures.
In addition, the critical wavelength is shorter than
that of the Rayleigh area-minimizing instability;
the polymer-mediated interaction may trigger the undulation instability.

When the bare elastic modulus of a membrane is not small,
for long wavelength undulations,
the effect softens the surface tension of the membrane box.
On the other hand, for short wavelength undulations,
the effect hardens the surface tension because of the
reduction of the entropy of confined polymers.
It is quite remarkable that the discrimination point of softening or
hardening (instabilizing or stabilizing) is determined by the distance
between membranes.

One should be careful to interpret our results.
First, because of the volume preserving condition,
the undulation is not a simple expansion of membrane boxes or tubes
by the thermal motion of polymers.
Second, the calculation of wave equations does not correspond to the case
when the wave length of undulations
is longer than confined polymer sizes in boxes or tubes.

The polymer-mediated entropic force without undulations is
proportional to $d^{-3}$, where $d$ is the distance
between walls, which decays slower than Van der Waals attractive interactions.
This force is something like the excluded volume interactions of
multimembranes systems known as the Helflich repulsive interaction.
Finally, the polymer confinement rule for excluded volume chains is modified as
$(\Delta x)^{5/3} \Delta S \sim k_{\rm B} R_{\rm g}^{5/3}$.
We expect that the same entropic effect will exist for excluded volume chains.

\section*{Appendix A}

To tackle the integration of Eqs.(\ref{A1})-(\ref{A2}),
we define an integral:
\begin{eqnarray}
<f(x)>=\frac{a}{2 \pi}\int_{-\frac{\pi}{a}}^{\frac{\pi}{a}}
\frac{f(x)\,dx}{1+\epsilon \sin ax}.
\end{eqnarray}
By using a formula,
\begin{eqnarray}
\int \frac{dx}{1+\epsilon \sin x} =
\frac{2}{\sqrt{1-\epsilon^2}} \tan ^{-1}
\left(
\frac{\tan \frac{x}{2} + \epsilon}{\sqrt{1-\epsilon^2}}
\right),
\end{eqnarray}
we have
\begin{eqnarray}
<1>= \frac{1}{\sqrt{1-\epsilon^2}}.
\end{eqnarray}
Hence, it yields
\begin{eqnarray}
<\sin ax> &=& -A/{\epsilon},
\\
<\sin^2 ax> &=& A / \epsilon^2,
\\
<\cos^2 ax> &=&  B / \epsilon^2,
\\
<\cos^2 ax \sin ax> &=& - C / \epsilon^3,
\\
<\cos^2 ax \sin^2 ax> &=&  C / \epsilon^4,
\end{eqnarray}
where $A, B$ and $C$ are defined in Eqs.(\ref{ABC})$-$(\ref{ABCD}).

\section*{Appendix B}
In this appendix, we provide some integrals of the Bessel functions.
By using
$J_0'(x) =-J_1(x)$ and $(xJ_1(x))'=xJ_0(x)$,
it is easy to verify the following integrals:
\begin{eqnarray}
\int dr rJ_0^2(c r)=[\frac{r^2}{2}(J_0^2(c r)+J_1^2(c r))],
\\
\int dr r^2J_0(c r)J_0'(c r)=
[-\frac{r^2}{2c}J_1^2(c r)],
\\
\int dr r^3J_0^2(c r)=
\frac{1}{6}[r^4J_0^2(c r)
+(r^4-\frac{2}{c^2}r^2)J_1^2(c r)
\nonumber
\\
+\frac{2}{c}
r^3J_0(c r)J_1(c r)].
\end{eqnarray}
Then, we obtain
\begin{eqnarray}
\int_0^R dr rJ_0^2\left(\frac{\alpha r}{R}\right)
&=&\frac{R^2 \beta}{2},
\\
\int_0^R dr r^2J_0\left(\frac{\alpha r}{R}\right)
J_0'\left(\frac{\alpha r}{R}\right)
&=&-\frac{R^3 \beta}{2\alpha},
\\
\int_0^R dr r^3J_0^2\left(\frac{\alpha r}{R}\right)
&=&\frac{R^4 \beta}{6}(1-\frac{2}{\alpha^2}).
\end{eqnarray}

%\section{NOTES}

%\begin{eqnarray*}
%\int_{-w/2}^{w/2}  y
%\sin \left(\frac{\pi y}{w}\right)
%\cos \left(\frac{\pi y}{w}\right)
%dy =\frac{w^2}{4\pi},
%\\
%\int_{-w/2}^{w/2}  y^2
%\cos^2 \left(\frac{\pi y}{w}\right) dy
%=\frac{w^3}{4}\left(\frac{1}{6}-\frac{1}{\pi^2}\right).
%\end{eqnarray*}

\end{multicols}

\begin{thebibliography}{99}
\bibitem{Ca}
E. F. Cassasa, J. Polymer Sci., B{\bf 5}, 773 (1967).
See also,
P. G. de Gennes,
{\it Scaling Concepts in Polymer Physics}
(Cornell University Press, Ithaca and London, 1979).

\bibitem{Daoud}
M. Daoud and P. G. de Gennes,
J. Phys. France, {\bf 38}, 85 (1977).

\bibitem{BC}
J. T. Brooks and M. E. Cates,
J. Chem. Phys., {\bf 99}, 5467 (1993).

\bibitem{Lal}
J. Lal, S. K. Sinha, and L. Auvray,
J. Phys. II France, {\bf 7}, 1597 (1997).

\bibitem{GJ}
J. Goldstone and R. L. Jaffe,
Phys. Rev. B {\bf 45}, 14100 (1992);
See for a review,
P. Duclos and P. Exner,
Reviews in Mathematical Physics, {\bf 7}, 73 (1995).

\bibitem{YPSW}
K. Yaman, P. Pincus, F. Solis, and T. A. Witten,
Macromolecules, {\bf 30}, 1173 (1997);
K. Yaman, P. Pincus, and C. M. Marques,
Phys. Rev. Lett. {\bf 78}, 4514 (1997);
K. Yaman, M. Jeng, P. Pincus, C. Jeppesen, and C. M. Marques,
Physica, A {\bf 247}, 159 (1997);

%\bibitem{SD}
%Y. Y. Suzuki, T. Dotera, and M. Hirabayashi, in preparation;
%T. Dotera and Y. Y. Suzuki, in preparation.

\bibitem{Safran}
S. A. Safran,
{\it Statistical Thermodynamics of Surfaces, Interfaces, and Membranes}
(Addison-Wesley, Reading, 1994).

\bibitem{DWT}
D. W. Thompson,
{\it On Growth and Form, the Complete Revised Edition} (Dover, New York, 1992).

\bibitem{Unpub}
P. G. de Gennes,
{\it Simple Views on Condensed Matter: Expanded Edition}
(World Scientific, Singapore, 1998), p.405.
Chain reflections on walls are known to cause
the surface segregation of chain ends.

\end{thebibliography}
\end{document}